# Size effects in the thin films of order - disorder ferroelectrics subject to the depolarization field

E.A. Eliseev*, M.D. Glinchuk**

Institute for Problems of Materials Science, National Academy of Science of Ukraine,

Krjijanovskogo 3, 03142 Kiev, Ukraine

*E-mail: eliseev@mail.i.com.ua

**E-mail: glin@materials.kiev.ua

**Abstract**

The films of order-disorder type ferroelectrics were considered in the mean field approximation taking into account depolarization field. It was shown that size effects in this system could be described on the base of bulk system equation of state with Curie temperature dependent on the film thickness.

The critical size $h_c$ and critical temperature $T_c$ of phase transition from ferroelectric to paraelectric phase was calculated allowing for the depolarization field contribution. The comparison of the polarization dependence on the film thickness, temperature and electric field for the films of order-disorder and displacement type ferroelectrics is performed. In particular it was shown that all the dipoles become ordered at $T \to 0$ independently on the film thickness for $h > h_c$ contrary to the displacement type ferroelectrics. Critical thickness appeared larger and polarization distribution sharper for the displacement type ferroelectrics than for order-disorder type ferroelectrics.

## 1 Introduction

Size effect drastically influences physical properties of thin ferroelectric films. It was shown recently [1], [2] for the displacement-type ferroelectrics that depolarization field significantly changes phase transition characteristics in thin ferroelectric films even in the short-circuited system. For example, depolarization field can destroy ordered phase at thickness much larger than critical one for the system where depolarization field is absent [1], [2].

However the existing model of size effects in the confined system of the order-disorder type ferroelectrics neglected the effects of depolarization field even for mono domain ferroelectric film polarized normally to its surface [3], [4], [5].

In this paper we use the simplest model of order-disorder type system without effects of tunneling and the four-dipoles interactions. However, we take into account the depolarization field, arising from uncompensated bound charges inside the inhomogeneously polarized ferroelectric films [6].

## 2 Basic equations

Let us consider the system of electric dipoles with two possible orientations. Hamiltonian of the system has the form:

$$H = -\frac{1}{2}\sum_{i,j} J_{ij} l_i l_j - \sum_i p E_i l_i \qquad (1)$$

Here $J_{ij}$ is potential of interaction between dipoles with unit vectors $l_i$ and $l_j$, $p$ is dipole moment, $E_i$ is the electric field acting on the dipole $l_i$. This field is the sum of the external field $E_{ext}$ and $E_d$ depolarization field. Summation in Eq. (1) on the indexes $i$ and $j$ is carried out on dipoles sites.

With the help of Hamiltonian (1) in the molecular field approximation [7] one can obtain the following equation for the thermally averaged vectors $\bar{l}_i$:

$$-\frac{1}{2}\sum_j J_{ij} \bar{l}_j - p E_i + k_B T \operatorname{arctanh}(\bar{l}_i) = 0 \qquad (2)$$

Here $\bar{l}_i$ determines thermally averaged dipole moment of dipole i or the order parameter distribution across the system. The equation of this type was considered earlier [5] for the transverse Ising model in the linear approximation and without electric field.

Using the continuous approximation $\bar{l}_i \equiv l(\vec{r}_i)$ and the relationship $l(\vec{r} \pm \vec{a}) = l(\vec{r}) \pm \vec{a}\nabla l(\vec{r}) + (\vec{a}\nabla l(\vec{r}))^2 \pm \ldots$ (see e.g. [3], [4]) it is easy to write the equation determining the spatial distribution of order parameter:

$$-J l(\vec{r}) - J \delta \Delta l(\vec{r}) - p(E_{ext} + E_d) + k_B T \operatorname{arctanh}(l(\vec{r})) = 0 \qquad (3)$$

Here $J$ is effective interaction constant or the mean field acting on the each dipole from its neighbors, $\delta$ determines the correlation between dipole moments. For Ising model $J$ is proportional to the interaction constant and number of nearest neighbors, $\delta$ is squared distance between dipoles [3]. It can be shown that for the interaction involving not only neighbor dipoles the quantity $\delta$ can be several orders larger than squared distance between dipoles.

For the bulk system one can suppose order parameter as homogeneous one, so depolarization field is absent because of the short-circuit conditions and Eq. (3) can be expressed in the well-known form of the equation of state for the bulk order-disorder type ferroelectrics in the mean field approximation $l_\infty = \tanh((pE_{ext} + J l_\infty)/k_B T)$. It is seen that in the absence of external field this equation has nonzero solution only at temperature below Curie temperature $J/k_B$.

Hereafter we consider ferroelectric film with thickness $h$ and dipoles perpendicular to the film surface. In this case depolarization field is not zero and differential equation (3) for the confined system has to be supplemented with the boundary conditions. These conditions depend on the surface model. The most general form of boundary condition introduces so-called extrapolation length $\lambda$. With axis $z$ perpendicular to the film surfaces these conditions can be written as:

$$\left(\lambda \frac{dl(z)}{dz} + l(z)\right)\bigg|_{z=h/2} = 0, \quad \left(\lambda \frac{dl(z)}{dz} - l(z)\right)\bigg|_{z=-h/2} = 0. \tag{4}$$

Hereafter we consider the case of the positive extrapolation length only.

The electric field E in the film between ideal superconducting electrodes with fixed electric potentials has the view [6], [1], [2]:

$$E(z) = E_{ext} + 4\pi np(\langle l \rangle - l(z)), \quad \langle l \rangle = \frac{1}{h}\int_{-h/2}^{h/2} l(z)dz. \tag{5}$$

Here $n$ is the concentration of dipoles, $n\,p\,l(z)$ determines the polarization of the system.

The solution of linearized Eq. (3) allowing for boundary conditions (4) and depolarization field (5) can be easily expressed with elementary function [6]. However, this solution corresponds to the paraelectric phase only. In ferroelectric phase non-linearity has to be taken into account. For Landau-Ginsburg-Devonshire type ferroelectrics this was done with the help of the direct variational method [1]. We choose the solution of linearized Eq. (3) as a trial function and an amplitude factor will be treated as a variational parameter:

$$l(z) = L(1 - \varphi(z)), \quad \varphi(z) = \frac{\cosh(z/l_d)}{\cosh(h/2l_d) + \sinh(h/2l_d)\lambda/l_d}. \tag{6}$$

Here $L$ is the variational parameter, which has the meaning of the order parameter maximal value, $l_d = \sqrt{J\delta/(4\pi np^2 + k_B T - J)}$ is characteristic length. It is easy to show that ratio $4\pi np^2/J$ is equal to the ratio of Curie-Weiss constant ($C_{CW}$) to the Curie temperature ($T_C$) in paraelectric phase of bulk system. So that $4\pi np^2/J \gg 1$ for the most of ferroelectrics and the length $l_d$ is always real and almost independent on temperature: $l_d \approx \sqrt{J\delta/4\pi np^2}$. Parameter $L$ can be found from the following equation:

$$\left(-1+\left(1+\frac{\delta}{l_d^2}\right)\Phi\right)L + \frac{k_B T}{J}\langle\text{arctanh}(L(1-\varphi(z)))\rangle = \frac{pE_{ext}}{J}, \quad \Phi = \frac{\tanh(h/2l_d)2l_d/h}{1+\tanh(h/2l_d)\lambda/l_d}. \quad (7)$$

This equation can be essentially simplified. Namely, for the films with thickness $h$ much larger than $l_d$ quantity $\Phi \approx 2l_d^2/(h(l_d+\lambda))$ is much smaller than unity and one can neglect the term $\varphi(z)$. Taking into account that $\delta/l_d^2 \gg 1$, Eq. (7) can be rewritten as follows:

$$\left(-1+\frac{2\delta}{h(l_d+\lambda)}\right)L + \frac{k_B T}{J}\text{arctanh}(L) = \frac{pE_{ext}}{J} \quad (8)$$

This equation can be rewritten in the following form

$$L = \tanh\left(\frac{pE_{ext}}{k_B T} + \frac{LJ}{k_B T}\left(1-\frac{h_c}{h}\right)\right), \quad h_c = \frac{2\delta}{\lambda+l_d}. \quad (9)$$

Here $h_c$ is a critical thickness of the film because $L\to 0$ at $h\to h_c$ without external field. For the films thinner then $h_c$ there is no order phase (at zero external field) for the arbitrary temperature.

It should be noted that the same expression for the critical thickness of the displacement type ferroelectric films was obtained earlier [1], [2]. This allows us to make some estimation for the different type ferroelectrics. As it was mentioned above the value $\delta \sim 10 \div 100 a^2$, where $a$ is the lattice constant, and $4\pi n p^2/J = C_{CW}/T_C \sim 10$ for order-disorder type ferroelectrics [8]. Within these limits one can easily obtain that $l_d \sim 1 \div 3a$. For the displacement type ferroelectrics $\delta$ value lies within the same limits (see, e.g. [9]), but $C_{CW}/T_C > 10^2$ [8] and $l_d \sim 0.3 \div 1a$. Using these values, we estimate critical thickness for the different extrapolation length values (see Table 1) with $\delta = 100 a^2$.

**Table** 1. Critical thickness for different type ferroelectrics.

| λ | $h_c$ | |
|---|---|---|
| | order-disorder type ferroelectrics ($l_d=3a$) | displacement type ferroelectrics ($l_d=a$) |
| $a$ | 50 $a$ | 100 $a$ |
| 10 $a$ | 15 $a$ | 20 $a$ |
| 30 $a$ | 6 $a$ | 6.5 $a$ |

It is seen from Table 1 that for displacement type ferroelectrics critical thickness is larger than for order-disorder type ferroelectrics for which $h_c$ is several times larger than $l_d$ at $\lambda \sim l_d$ so that Eq. (9) is valid for $h \geq h_c$.

The equation of state (9) describes phase transition with critical temperature dependent on film thickness:

$$T_{cr} = \frac{J}{k_B}\left(1 - \frac{h_c}{h}\right), \quad h_c \geq h. \tag{10a}$$

It is clear that with the film thickness increase the transition temperature tends to Curie temperature of the bulk system. Furthermore, equation (9) determines critical thickness at which size driven phase transition occurs for the fixed nonzero temperature:

$$h_{cr} = h_c \frac{J}{J - k_B T}, \quad k_B T \leq J. \tag{10b}$$

It is obvious that this critical thickness $h_{cr}$ is always larger than value $h_c = h_{cr}(T=0)$.

These results for the order-disorder type ferroelectric film exactly coincide with obtained recently phase transition characteristics of the displacement type transition [1], [2]. It is because in the vicinity of the second order phase transition order parameter tends to zero and different type ferroelectrics are indistinguishable. However, the difference can be seen at $T<T_{cr}$ in the details of the properties behaviour. In addition, values of the length $l_d$ determining order parameter spatial distribution for displacement type ferroelectrics is several times smaller than for order-disorder type ferroelectrics. This means that in the first case spatial distribution is more abrupt than in the second case.

Order parameter calculated with the help of Eqs. (6) and (9) as the function of temperature and external field is represented in Fig. 1a, b respectively for the different value of the film thickness above critical one. Curve 4 corresponds to the bulk material properties. It is seen that with the film thickness decrease transition temperature and coercive field also decrease. Order parameter as the function of film thickness and external field is represented in Fig. 2a, b respectively for the different value of the temperature below bulk material Curie temperature. As the temperature increases, critical thickness also increases but coercive field decreases.

It is seen from Fig. 1a that order parameter saturate to unity with temperature decrease even for the thin films in contrast to displacement type ferroelectric film [1], [2]. Another distinct feature of order-disorder films is fast saturation with the external field increase (see Fig. 1b and 2b).

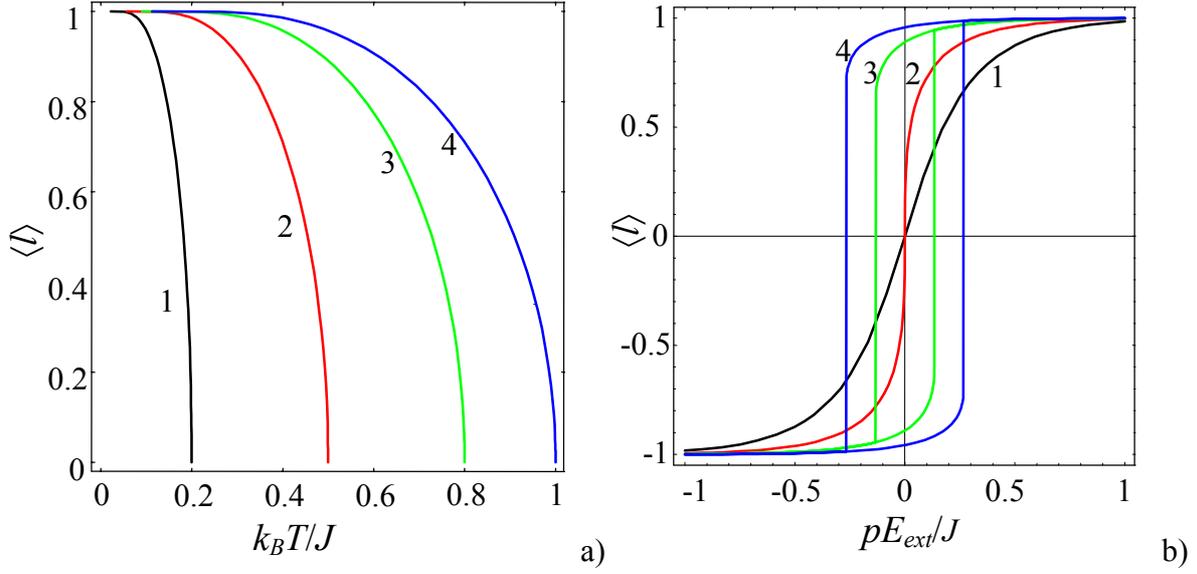

**Fig. 1** Averaged value of *l(z)* dependence on temperature at $E_{ext}=0$ (a) and on external field at $k_BT/J=0.5$ (b) for $h/h_c=1.25, 2, 5, \infty$ (curves 1, 2, 3, 4).

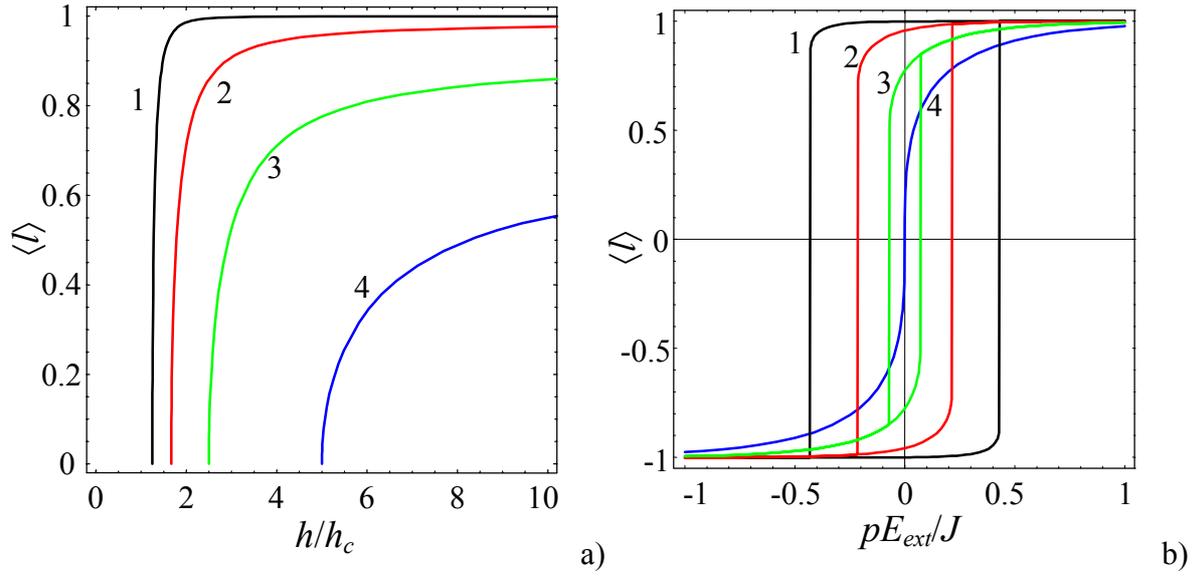

**Fig. 2** Averaged value of *l(z)* dependence on film thickness at $E_{ext}=0$ (a) and external field at $h/h_c=5$ (b) for $k_BT/J=0.2, 0.4, 0.6, 0.8$ (curves 1, 2, 3, 4).

It should be noted, that in the region $h<h_c$ Eq. (9) gives only qualitative description in comparison to the exact solution of linearized Eq. (3), which represents the linear dielectric susceptibility $\chi = n p (dl/dE_{ext})|_{E_{ext}=0}$ in paraelectric (disordered) phase where $l\neq 0$ at $E_{ext}\neq 0$:

$$\chi_{PE} = \frac{n p^2 (1-\varphi(z))}{(k_BT-J)(1-\Phi)+J\Phi\delta/l_d^2}, \quad \langle\chi_{PE}\rangle = \frac{n p^2}{k_BT-J\left(1-\dfrac{2\delta}{h(l_d \coth(h/2l_d)+\lambda)-2l_d^2}\right)}. \quad (11)$$

The critical thickness as well as the transition temperature determined from Eq. (11) is close to that from equation of state (9) both for small and great values of extrapolation length.

### 3 Conclusion

The equation of state (9) is similar to the one of the bulk system with Curie temperature dependent on the film thickness. The similar result was obtained for the displacement type ferroelectrics [1], [2] in the Landau-Ginsburg-Devonshire (LGD) theory framework. Really, one can expand the obtained equation of state (9) or initial equation (3) in powers of $l(z)$ and obtain LGD-type equation of state (see e.g. [3], [4]). However, this expansion valid only in the vicinity of the phase transition and all its nonlinear coefficients appreciably depend on the temperature. Besides it is impossible to cut off this expansion in the vicinity of $l \to \pm 1$ because formally it is divergent in this case.

It should be noted that effects of tunneling and the four-dipoles interactions can be easily taken into account, as well as effects of internal mechanical strain, arising due misfit between film and its substrate. These effects consideration is in progress now.